\documentstyle[aps,epsf]{revtex}
\draft
\title{Diffusion controlled initial recombination}
\author{T. Christen}
\address{ABB Corporate Research, CH-5405 Baden-D\"attwil, Switzerland}
\author{M. B\"uttiker}
\address{D\'epartement de physique th\'eorique,
Universit\'ee de Gen\`eve\\ 24, Quai Ernest-Ansermet, CH-1211 Gen\`eve,
Switzerland}
\twocolumn

\begin{document}
\maketitle
\begin{abstract}
This work addresses nucleation rates in systems with strong
initial recombination. Initial (or `geminate') recombination is a process
where a dissociated structure (anion, vortex, kink etc.) recombines with
its twin brother (cation, anti-vortex, anti-kink) generated in the same
nucleation event. Initial recombination is important if
there is an asymptotically vanishing interaction
force instead of a generic saddle-type activation barrier. At low
temperatures, initial recombination strongly dominates homogeneous
recombination.
In a first part, we discuss the effect in one-, two-, and three-dimensional
diffusion controlled systems with spherical symmetry. Since there is no
well-defined saddle, we introduce a threshold which is to some extent
arbitrary but which is restricted by physically reasonable
conditions. We show that the dependence of the nucleation rate on the
specific choice of this threshold is strongest for one-dimensional
systems and decreases in higher dimensions. We discuss also the influence
of a weak driving force and show that the transport current is
directly determined by the imbalance of the activation rate in
the direction of the field and the rate against this direction.
In a second part, we apply the results to the overdamped
sine-Gordon system at equilibrium. It turns out that
diffusive initial recombination is the essential mechanism which
governs the equilibrium kink nucleation rate. We emphasize
analogies between the single particle problem with initial recombination
and the multi-dimensional kink-antikink nucleation problem.
\end{abstract}
\narrowtext

\section{Introduction}
\label{Introduction}
The purpose of this work is to support our approach
to the nucleation problem of kinks and antik-kinks
in the overdamped sine-Gordon chain at equilibrium \cite{BUC95}
by investigating closely related problems
which are dominated by initial (or `geminate') recombination
processes. Initial recombination is a process
where a dissociated structure (anion, vortex, kink etc.) recombines with
its twin brother (cation, anti-vortex, anti-kink) generated in the same
nucleation event \cite{ONS38,CAL83,CAU77,DON91,HAB73}.
Our theory of kink-antikink nucleation \cite{BUC95} leads in the
thermal equilibrium state of the sine-Gordon chain
to an activation rate proportional to $\exp(-2E_{0}/kT)$
with an activation energy of twice the equilibrium kink energy $E_{0}$.
This can be compared with earlier work \cite{HMS88} which predicts
a rate proportional to $\exp(-3E_{0}/kT)$.
This significant difference in the nucleation rates
is a consequence of initial recombination of kink antikink pairs.
Although our discussion presents a clear physical picture,
our results are apparently not obvious
and have already given rise to discussions in the literature \cite
{HM96,BUC96}.
It is therefore necessary to further support and explain more deeply the
point of view and the approach taken in Ref. \cite{BUC95}.\\
Usually, nucleation theory \cite{KRA40} (for reviews, see
\cite{HTB90,MEL91})
is associated with the decay of a metastable state
across an activation barrier. A simple picture is a
Brownian particle which has to overcome a barrier
in order to leave the region of attraction of a potential well
(see Fig. \ref{fig1}a). One of the main tasks of nucleation theory is the
evaluation of the current flow out of the well for a given density
of particles in the well. The relevant time scale of
such an activated process is dominated by an
inverse Arrhenius factor $\exp (E_{0}/kT)$, where $E_{0}$, $k$,
and $T$ denote the activation energy, the Boltzmann constant,
and the temperature, respectively. The nucleation rate is then proportional
to the Arrhenius factor. The proportionality constant
is in the center of interest in many works and depends strongly
on the shape of the potential and on the strength of the damping
of the Brownian particle. In the present work we emphasize
the particularly interesting problem of nucleation and recombination
in systems with an asymptotically `flat saddle'
(see Fig. \ref{fig1}b). We will consider activation
out of a well onto a region where
the potential converges fast to a constant value, without
crossing a barrier with a maximum. Except for a few remarks on
the underdamped limit we will restrict
ourselves to strongly damped systems. This corresponds to a diffusion in
configuration space on the flat saddle region. Furthermore we will allow
a very slow drift due to a weak external driving force.\\
The main
difference to the usual nucleation across a saddle with a well
defined maximum consists in an enhanced probability of backscattering
into the well. Indeed, while there the particle is driven away by
the deterministic force once the barrier maximum is traversed,
the particle on a flat saddle executes a diffusive motion leading to a
large probability of falling back into the original well.
This process corresponds to initial recombination in
generation-recombination kinetics and is in contrast to homogeneous
recombination. The homogeneous recombination rate is
proportional to the density of wells, and is much smaller than the rate of
initial recombination for a diluted well density.
The processes are illustrated in the lower part of  Fig.\ref{fig1}. \\
\indent
In the following we mention some examples of initial recombination.
The most prominent one appears in dissociation theory of diffusion
controlled  chemical reactions \cite{CAL83}.
In his 1938 paper titled `initial recombination of ions' \cite{ONS38},
Onsager determined the probability of recombination of a pair of ions
after a given initial separation $r$. It is well-known, that despite
the long-range nature of the Coulomb force,
there is a finite escape probability in three dimensions even at zero
external driving force. Onsager gives for the probability of escape
$\exp(-r_{o}/r)$, where $r_{o}=q_{1}q_{2}/(4\pi \epsilon _{r}
\epsilon _{0} kT)$ is called the Onsager radius, $q_{1,2}$ denote the
charges of anion and cation, and $\epsilon _{r} \epsilon _{0}$ is the
dielectric permeability of the medium. While for small initial distances
$r$ the attractive Coulomb interaction leads to strong initial
recombination,
diffusion dominates Coulomb interaction for $r$ larger than
the Onsager radius which leads to a large escape probability.
It is clear that in a dilute electrolyte and if the dissociation
process is due to thermal nucleation, initial recombination dominates
homogeneous recombination. This has strong consequences on the free
ion density in a superimposed electric field, since initial
recombination
is very sensitive to an external force. Note that in a flat potential,
i.e.,
if $r_{o}=0$, Onsager's result predicts a vanishing initial recombination
rate.
We will show below, that this is characteristic for a three-dimensional
system,
and it will be different for lower dimensions.\\ \indent
A two-dimensional example is the two-dimensional Coulomb
gas. Here, the force vanishes proportional to $1/r$ which leads to a
logarithmically increasing potential. Though the potential is not `flat',
this case is particularly interesting (see, e.g., \cite{MIN87}).
Dissociation in the 2-d Coulomb gas describes,
for instance, vortex-antivortex nucleation in superfluid Helium films.
An important feature of this system is that there is an unbinding
transition
of pairs at a critical temperature. At low temperatures
the escape probability vanishes,
whereas above a critical temperature it is finite. This transition, which
is known as the Kosterlitz-Thouless transition, can be seen as a
consequence of
strong initial recombination which inhibits free excitations
below the critical temperature \cite{CAU77}.
We mention, that there is no simple single-particle picture of the
Kosterlitz-Thouless transition since screening of the excitations
play a major role.\\ \indent
Initial recombination in one-dimensional systems is faced, e.g.,
in photoelectric carrier generation in 1-d semiconducting
polymers \cite{DON91}. Haberkorn and Michel-Beyerle \cite{HAB73}
considered electrons which are photo-generated close to an
electrode in a one-dimensional conductor,
including the image-force at the electrode. In particular, they discussed
the relation between the total current due to a weak external force, and
the `nucleated' current. We will take up these results below,\\  \indent
Note that the listed examples are all controlled by an interaction force
rather than
by diffusion alone. In this paper, however, we emphasize diffusion
controlled,
i.e., entropy driven, initial recombination. This requires that the force
be short ranged. Short range forces occur, e.g., between
kinks or domain walls in (quasi) one-dimensional
systems. A kink-antikink interaction potential is often a
monotonously increasing function of the separation and becomes
exponentially flat
for large separations \cite{KAW82}.
\\ \indent
The paper is organized as follows. In order to fix the notation
and for later comparison, we briefly review in
Sect. \ref{nuc} some standard results of the theory of nucleation
across a barrier. In Sect. \ref{flat} we discuss the nucleation
across flat saddles, including the effect of a
weak force, and the initial recombination time.
In Sect. \ref{kink}, diffusion controlled initial recombination
is illustrated for the example of equilibrium kink nucleation.

\section{Escape across a barrier}
\label{nuc}
In the following we consider the motion of a particle in $d$ dimensions
with coordinate $\vec r$, mass $m$, damping constant $\gamma $, and
in a potential $U(\vec r)$. Of course, the particle coordinate
vicariously stands for relevant variables of a rather large class of
systems
(chemical reaction variables, ion separation distance, liquid droplet
radius, magnetization, etc.).
Throughout this paper, we assume that the potential $U$ has a
minimum at $\vec r=0$ and depends only on the distance $r= \vert \vec r
\vert$. In a one-dimensional system ($d=1$), this means that $U(r)$ is an
even
function of the coordinate $r$. The particle is
furthermore coupled to a heat bath of temperature $T$. The equation of
motion for $\vec r$ is a Langevin equation with a white noise
force \cite{RIS84}. Equivalently, the system can be described by
Kramers equation for
the phase-space probability density $f(\vec r, \vec v,t)$, where
$\vec v$ is the velocity of the particle (for our purposes we may use
the velocity instead of the momentum). Below, we consider almost always
the strongly damped case in which Kramers equation reduces to the
Smoluchowski equation,
\begin{equation}
\partial _{t} P+ \nabla \vec j = 0 \;\;,
\label{smoluchowski}
\end{equation}
which is a continuity equation for the probability density
$P=\int d^{d}v\, f$ in configuration space. The probability current
$j=\int d^{d}v\, vf$ is given by
\begin{equation}
\vec j = -\frac{1}{m\gamma} \left(P\nabla U +kT \nabla P \right) \;\;.
\label{procurrent}
\end{equation}
where the Einstein relation $D=kT/m\gamma $ between diffusion
constant $D$ and damping constant $\gamma $ is related to local
equilibrium.\\ \indent
The nucleation rate $J_{nuc}$ can be defined as follows. First,
one has to find a stationary solution of
Kramers equation which is normalized in the well. Outside the well,
this solution has to satisfy appropriate boundary conditions. Usually,
absorbing boundary conditions are applied. Then, the
current crossing the boundary is the nucleation rate.
This approach has the advantage that the determination
of the nucleation rate is reduced to the solution
of a stationary (time-independent) problem.\\ \indent
In the remaining part of this section, we recall some results for
one-dimensional
systems and for saddles which are not flat in the above mentioned sense.
Furthermore, we restrict the discussion to low temperatures, i.e.,
$kT\ll E_{0}$. For moderate-to-strong damping, Kramers found for a
quadratic saddle
at $r=s$ an analytical result which reduces in the strong
damping limit to the Smoluchowski rate
\begin{equation}
J_{nuc} = 2\; \frac{\omega _{0} \omega _{S}} {2 \pi
\gamma}\exp{(-E_{0}/kT)} \;\;.
\label{rate1}
\end{equation}
The rate depends on the Arrhenius factor $\exp{(-E_{0}/kT)}$,
on the damping constant $\gamma $, and on the curvatures at the minimum and
 at the
saddle, $ \omega _{0} = \sqrt{\partial _{r}^{2} U(0) /m}$
and $ \omega _{S} = \sqrt{\vert \partial _{r}^{2} U(s)\vert /m}$,
respectively. The factor $2$ in Eq. (\ref{rate1}) occurs because
the symmetry of the one-dimensional
potential implies two equivalent paths of escape from the well.
In the weak damping limit ($\gamma \to 0$), Kramers moderate-to-strong
damping result yields the transition state rate
\begin{equation}
J_{TST} =  2 \; \frac{\omega _{0}} {2 \pi }\exp{(-E_{0}/kT)} \;\;.
\label{rate2}
\end{equation}
This result is wrong since thermal equilibrium is
assumed in the well which is no longer the case at very low
damping. An appropriate (slow) variable is then the action variable
$I$ (or the energy $E$). By an averaging over the fast angle
variable, Kramers derived therefore a diffusion equation
in action space. A solution of the nucleation problem in this
space yields the true rate in the $\gamma \to 0$ limit:
\begin{equation}
J_{nuc} = \gamma \frac{\omega _{0}I_{S}} {2 \pi kT}\exp{(-E_{0}/kT)} \;\;.
\label{rate3}
\end{equation}
Here $I_{S}$ is the action of the separatrix trajectory at the saddle
in phase space. We mention, that the result (\ref{rate3})
remains valid also for the `flat saddle' in Fig. \ref{fig1}b,
since at $\gamma \to 0$ the particle escapes ballistically,
i.e., without backscattering in the region with constant
potential.\\ \indent
The rates (\ref{rate1}) and (\ref{rate3}) are the leading order
results in $1/\gamma $ and $\gamma$, respectively.
The crossover between these limits is known as the
Kramers turnover problem.
Renewed attention to this problem followed Refs. \cite{BHL83,BUT88}
which investigated the leading corrections to Eq. (5) and provided
computational results illustrating the turnover.
A particularly elegant approach to this problem was put forth
by Melnikov \cite{MEL91,MEL86}, and independently by Risken et al.
\cite{RIS85,RIS88} based on boundary layer theory.
Later it has been noticed that the problem
can be solved over the entire turnover
region by normal mode analysis\cite{POL89}.
We expect that for the `flat saddle',
the first correction term to Eq. (\ref{rate3}) is strongly changed as
compared to the quadratic saddle. In particular, there should be
a strong dependence on the length of the diffusion region. In
the present work, however, we consider the strongly
damped case, or corrections to this case for finite but small
$1/\gamma$. To find the corrections, we will use a result of
boundary layer theory.\\ \indent

\section{Escape across a flat potential region}
\label{flat}
We consider again a symmetrical potential well. For decreasing $r < \xi_{0}
 $,
the potential is assumed to drop strongly and monotonously to its minimum
value $-E_{0}$ at $r=0$. For $r > \xi_{0} $, it is assumed to
increase monotonously and to converge fast to zero.
Here $\xi_{0} $ is a characteristic half-width
of the well defined by the specific physical problem under consideration
(see, e.g., Sect.\ref{kink}). A typical example is sketched in Fig.
\ref{fig1}b.
As mentioned above, for the one-dimensional case $d=1$,
we assume reflection symmetry, and in higher dimensions we
assume a rotationally invariant potential, which allows us to confine
all calculations to $r\geq 0$. On a large length scale
$L$ ($\gg \xi_{0}$), we assume a periodic lattice of such wells.
A two-dimensional sketch is shown in Fig. \ref{fig1}c. We define the
occurrence of a nucleation event as the escape of the particle from the
region $ r  < s $.
Initial recombination denotes recombination with the original well,
while we define here homogeneous recombination as capture by a neighboring
well (Fig. \ref{fig1}d). We emphasize that homogeneous recombination
and initial recombination differentiate histories of particles
with separations larger than $s$. A homogeneous recombination event is
thus represented by a trajectory that starts from $s$ and ends up at a
different
well, whereas initial recombination is represented by a trajectory
that starts from $s$ and after excursion in the flat potential
region returns to the initial well. Initial recombination refers thus
like homogenous recombination to particles that have completely
dissociated.
The only difference is their different
subsequent history due to diffusive motion.
\\ \indent
In order to determine the nucleation rate, we consider first a single well.
The bare potential $U(r)$ does not exhibit a saddle
and the particle
feels an attracting force which, however, becomes
(exponentially) small with increasing
distance from the well.
Once the particle is far enough away, it behaves thus purely diffusive and
must be considered to be free. As in standard nucleation theory, the
solution
of the Kramers problem must obey absorbing boundary conditions
outside the well. The introduction of the point $r=s$ on the
flat region where these boundary
conditions are applied defines the size of a fictitious saddle
separating the bound
state from the free states. The exact location of this point is
not determined due to local translational invariance in the region
of constant potential. However, besides satisfying
$L\gg s\gg \xi_{0}$, it should be fixed by physical conditions.
A typical example of an experimental condition which provides a certain
value
of $s$ is the finite resolution of an instrument which counts the free
particles \cite{BUC96}.\\ \indent
To calculate the nucleation rate, we solve the stationary
Smoluchowski Eq. (\ref{smoluchowski}) in radial coordinates.
Due to rotational symmetry there is
no dependence on angle coordinates for $d>1$.
With an ansatz $P(r)=\beta (r)\exp (-U(r)/kT)$
for the probability density, one finds from
Eq. (\ref{smoluchowski}) for the radial current density in $d$ dimensions
\begin{equation}
j_{r} = -D \exp (-U(r)/kT) \partial _{r} \beta (r) =
\frac{C}{r^{d-1}}\;\;.
\label{reduced}
\end{equation}
Here, $C$ has to be determined and is related to the nucleation rate
by $J_{nuc}= a_{d}C$ with $a_{1}=2$, $a_{2}=2 \pi$, and
$a_{3}=4 \pi$. The density is assumed to be normalized
\begin{equation}
\int_{0} ^{s} P(r)\, a_{d}r^{d-1}dr= 1
\label{normalization}
\end{equation}
and to satisfy the linear, homogeneous, and mixed boundary conditions
at $s$ \cite{BUR81}
\begin{equation}
x_{m}\partial _{r} P|_{s}+P|_{s}=0 \;\;.
\label{boundary}
\end{equation}
Some comments concerning this boundary condition are in order
(see, e.g., \cite{CAL83}). First, $x_{m}=\alpha \sqrt{D/\gamma}$
is the Milnes length which is a measure for the nonequilibrium
boundary layer at an absorbing boundary. Indeed,
$x_{m}\propto \sqrt{kT/m}/\gamma$ can be interpreted as the mean
free path of a particle with thermal velocity. In one dimension,
$\alpha = \zeta (-1/2) \approx 1.46$. In higher dimensions, $\alpha $
is changed (e.g., due to the finite curvature of the boundary),
but here we are not interested in details concerning this problem
and we will use $\alpha $ as a given parameter. In the limit $\gamma
\to \infty$, an absorbing boundary at $s$ implies $P(s)=0$. On a
microscopic
length scale, the boundary condition (\ref{boundary}) describes
the solution asymptotically far away from the
absorbing boundary, $r< s-x_{m}$. The boundary layer itself ($s-x_{m}<r<s$)
shows a more complicated structure \cite{BUR81,HAR81}. Hence, Eq.
(\ref{boundary}) cannot be used if $ x_{m}$ is of the order of $s$,
or if the boundary layer leaks into the well.
The mixed boundary condition (\ref{boundary}) serves only to investigate
how the results presented below change as we depart from
the regime of strongly damped motion; it should be clear
that for weak friction,
it is not sufficient to consider the Smoluchowski
equation but instead one has rather to use the full
Kramers equation. \\ \indent
Let us now continue with the derivation of the nucleation rate.
Integration of (\ref{reduced}) together with (\ref{normalization}) and
(\ref{boundary}) yields
\begin{equation}
J_{nuc}=D\left( \int _{0} ^{s} r^{d-1}\, dr\, e^{-U(r)/kT}
\left( \frac{x_{m}s^{1-d}}{e^{-U(s)/kT}}+\int _{r}^{s}\frac{z^{1-d}\,  dz}{
e^{-U(z)/kT}} \right) \right)^{-1}
\label{solution1}
\end{equation}
Alternatively, we could derive this result by evaluating
the mean first passage time (see, e.g., \cite{HTB90}) of a particle
starting at $r=0$ and reaching $r=s$, with appropriate
boundary conditions.\\ \indent
Despite  rotational symmetry of the saddle manifold for $d>1$,
the rate cannot be determined
by using the volume of the symmetry group in the
order-parameter ($\vec r$) space \cite{LAN69}.
This works only for a saddle with a well defined sharp
maximum. We anticipate that the nucleation rate
(\ref{solution1}) depends in general on $s$.
The discussion of the dependence of the nucleation rate on this parameter
is our main goal. In the sequel the high
and low temperature limits will be discussed separately.

\subsection{High-temperature limit}
For very high temperatures $kT\gg E_{0}$, the particle
sees an overall flat potential and does not feel the well.
One finds from (\ref{solution1}) a rate
\begin{equation}
J_{nuc}=\frac{D}{s^2}\, \frac{2d}{1+2(x_{m}/s)}\;\;.
\label{solution2}
\end{equation}
In the high-friction limit this describes just the spatial diffusion out of
a region $r<s$. In the low-friction limit, on the other hand,
the rate is $dD/(sx_{m})=\sqrt{kT/m}/s\alpha$. Although this result
cannot be right as discussed above, let us compare it
with an estimate of the zero-friction result obtained
directly from phase-space considerations. In the limit $x_{m}\gg s$,
the probability density in phase space, $f(r,v)$,
must be considered rather than the configuration space density
$P(r)$.
For a constant potential, we take $f=A\exp(-mv^{2}/2kT)$ for radial
velocities pointing away from the well, and $f=0$ for velocities pointing
towards the well. In the following, we restrict ourselves to the
one-dimensional case. The density in configuration space ($r\geq 0$)
and the current density
are $P=\int dv f= A\sqrt{\pi kT/2m}$ and $j=\int vdv f= kTA/m$,
respectively.
Normalization implies $P=1/2s$ and we find
$J_{nuc}=2j=(\sqrt{2/\pi})(\sqrt{kT/m})/s$.
Up to a constant factor of order one this is in rather good agreement
with the result derived from Eq.(\ref{solution2}).\\ \indent
Let us briefly determine the recombination time $\tau_{1}$ in the
overdamped one-dimensional case. Due to equipartition at high temperatures,
 the ratio
of free particles ($r>s$) to particles which are not free ($r<s$)
is given by $N_{\rm free}/N_{\rm well}=(L-s)/s\approx L/s$.
Now, the total nucleation rate and the recombination rate balance
each other, $N_{\rm well}J_{nuc}=N_{\rm free}/\tau_{1}$. Using
Eq. (\ref{solution2}), this immediately
yields the recombination time $\tau _{1}= Ls/2D$. In part C of this
section,
we will discuss the recombination times more deeply.\\ \indent
\vskip 1.5cm
We notice that in the high temperature limit
the assumption of a diluted gas of nucleated structures,
i.e. the assumption $L\gg \xi_{0}$, breaks
down in many physical applications. This is because $L^{-d}$ is often a
density
of nucleated structures which is itself proportional to an Arrhenius
factor, and
becomes large at high temperatures. In that case, initial recombination
and homogenous recombination can no longer be clearly separated.

\subsection{Low-temperature limit}
In this case, the d-dimensional probability density
in the well is a normalized (see Eq. (\ref{normalization}))
equilibrium distribution
\begin{equation}
P_{0}(r) = A_{d} \exp(-U(r)/kT)
\label{ed}
\end{equation}
with the normalization constant
\begin{equation}
A_{d}= \left({\frac{m\omega _{0}^{2}}{2\pi
kT}}\right)^{d/2}\exp(-E_{0}/kT) \;\;.
\label{normconstant}
\end{equation}
The low-temperature nucleation rate is given by
\begin{equation}
J_{nuc}=a_{d}A_{d} D \left(\int _{\xi_{0} }
^{s} r^{1-d}\, dr + x_{m} s^{1-d} \right)^{-1}
\label{rateg}
\end{equation}
where $\xi_{0} $ is the size of the well.
It is interesting, that in one space dimension ($d=1$), the result
(\ref{rateg}) can be written in the form
\begin{equation}
J_{nuc}=\frac{\sqrt{2\pi}}{\alpha }\, \frac{x_{m}}{s-\xi_{0} + x_{m}}
J_{TST} \;\; ,
\label{rate1d}
\end{equation}
where $J_{TST}$ is the transition state rate
(\ref{rate2}). Note that despite of the vanishing of $x_{m}$
in the limit $m \to 0$, where the particle
is overdamped, the rate remains finite since $J_{TST}$
diverges. In the overdamped
limit and for $\xi_{0} \ll s$, the rate is proporional to $1/s$
and can be expressed in the form:
\begin{equation}
J_{nuc}=2 \frac{\omega _{0}}{\gamma s}\sqrt{\frac{kT}{2\pi m } }\exp
(-E_{0}/kT)  \;\; .
\label{rate1e}
\end{equation}
Equation (\ref{rate1e}) is the central result of this work.
According to Eq. (\ref{rate1e}), the characteristic decay time scales
linearly with $s$. In Sect. \ref{kink} we show that the equilibrium
nucleation reate of kinks has the same dependence on $s$.\\ \indent
For weak damping $x_{m}\gg s$,
we find that $J_{nuc}$ deviates from the transition state result
only by a factor of $1.73$. This (relatively small) deviation
is due to the above mentioned fact that the inner structure of the
boundary layer must be taken into account and cannot be
described simply by a boundary condition. One finds exactly the transition
state
result by an appropriate matching of the probability density $P=A\sqrt{\pi
kT/2m}$ (see end of the last subsection) to the part of the solution
(\ref{ed})
associated with right-moving particles ($r>0$). Hence, our result has a
behavior
similar to Kramers' moderate-to-strong damping formula for the quadratic
saddle. As mentioned, (\ref{ed}) is a wrong distribution at low damping
since local equilibrium in the well is not established, and
the nucleation rate at $\gamma \to 0 $ is given by Eq. (\ref{rate3}).\\
\indent
In two and three space dimensions, the rates can be expressed in the form
\begin{equation}
J_{nuc}=\frac{\omega _{0}^{2} }{\gamma }\, \frac{\exp(-E_{0}/kT)}
{\ln (s/\xi_{0})+x_{m}/s}
\label{rate2d}
\end{equation}
and
\begin{equation}
J_{nuc}=\frac{\omega _{0}^{2} }{\gamma }\,
\sqrt{\frac{m\omega _{0}^{2}}{2\pi kT}}
\frac{2\xi_{0}}{1-(\xi_{0} /s)+(\xi_{0} x_{m}/s^2)}
\exp(-E_{0}/kT)\;\; ,
\label{rate3d}
\end{equation}
respectively. In contrast to the one-dimensional case, the well size
$\xi_{0} $ cannot be neglected for $d>1$. Moreover, in two dimensions the
$s$-dependence is logarithmically weak, and in three dimensions its
influence can be neglected. We mention that a discussion of the result
in the limit $\xi_{0} \to 0$ requires some care, since $\xi_{0} $, $\omega
_{0}$, and $E_{0}$ are usually not fully independent.\\
Furthermore, in $d=2,3$, neither $\xi_{0} $ nor $s$ are necessarily equal
to the usual
transition state obtained by extremalization of the effective potential
$kT\ln(r^{1-d}\exp(U/kT)) = U(r) - kT(d-1)\ln(r)$. In contrast to the
usual quadratic saddle, it is not only the vicinity of this transition
state
which contributes to the integral in Eq. (\ref{solution1}). It is a larger
region around the transition state that counts, of which $\xi_{0} $ and
$s$,
respectively, are the lower and the upper boundaries of inegration.

\subsection{Lifetime of free particles}
\label{lfp}
We return now to the periodic lattice of
the wells sketched in Fig.\ref{fig1}c.
Each well can then be associated with
a unit cell of volume $V_{d}=L^{d}$ much larger than the volume
associated with bound particles ($\propto s^{d}$).
At equilibrium and at low temperatures, $kT\ll E_{0}$, the ratio of
particles in the well to free particles
is given by $N_{\rm well} /N_{\rm free}\approx 1/V_{d}A_{d}$.
In steady state, the generation rate $N_{\rm well}J_{nuc}$ equals the
homogeneous
recombination rate $N_{\rm free}/\tau $. The lifetime of
free particles is thus given by
\begin{equation}
\tau_{d} = \frac{V_{d}A_{d}}{J_{nuc}} \;\;.
\label{lifetime}
\end{equation}
For very strong damping and for $\xi_{0} \ll s$, Eqs.
(\ref{rate1d})-(\ref{rate3d}) imply
\begin{eqnarray}
\tau _{1} & = & \frac{Ls}{2D} \;\; ,\label{tau1} \\
\tau _{2} & = & \frac{L^{2}}{2\pi D}
\ln \left(\frac{s}{\xi_{0}}\right)\;\; ,
\label{tau2} \\
\tau _{3} & = & \frac{L^{3}}{4\pi D \xi_{0} } \;\; . \label{tau3}
\end{eqnarray}
In one dimension, the lifetime scales linearly with $s$, while the
$s$ dependence in higher dimensions is weak or negligible.
The higher the dimension the more unlikely is the
probability to reach the original well by pure diffusion. This is related
to a divergence of the lifetimes as $\xi_{0}$ vanishes. We also mention
that (\ref{tau3}) corresponds to the standard result in
the theory of diffusion controlled reactions, which states that the
rate per volume is given by $4\pi D \xi _{0}$
(see, e.g., \cite{CAL83}).\\ \indent
For the one-dimensional case, the lifetime can be calculated directly by
solving the Smoluchowski equation (\ref{smoluchowski}) for a flat
potential,
for absorbing boundary conditions $P(0)=P(L)=0$, and for an injected
current $j(s+0)-j(s-0)= j_{nuc}$ at $r=s$. If $P(r)$ is normalized,
the lifetime is then given by $1/j_{nuc}$. One finds that recombination
can be understood as a sum of two contributions, i.e. it holds
$j_{nuc}=j_{s}+j_{h}$, where
$j_{s}^{-1}= Ls/2D$ associated with (\ref{tau1}) and
$j_{h}^{-1}= L(L-s)/2D$ associated with diffusion to the neighbor well.\\
\indent
Below, in the discussion of equilibrium kink nucleation,
we will first calculate the lifetime independently from the nucleation
rate and from equilibrium statistical mechanics.
Using then both results for the nucleation rate and the lifetime,
we will derive the equilibrium kink density and show agreement with the
statistical mechanics result for the kink density.

\subsection{Effect of a weak force}
\label{force}
In this section we discuss the influence of an external weak force $F$ on
nucleation in one dimension and at low temperature.
The potential $U(r)$ is now replaced by the new
potential $U(r)-Fr$. We are interested in the response to leading order
with respect to $F$.
The saddle is still flat enough such that
diffusion dominates the drift in the relevant region: $kT\gg Fs$.
For finite $F$, the current to the left ($j_{-}$) and to the right
($j_{+}$) are no longer equal and the nucleation rate is
$J_{nuc}=j_{+}+j_{-}$. Interestingly, as we will show below,
the transport current $J_{trans}$ is determined by the
imbalance of the forward
nucleation $j_{+}$ and the backward nucleation
$j_{-}$. To leading order in $F$, the transport current is
thus given by
\begin{equation}
J_{trans}=j_{+} - j_{-} \;\;.
\label{trans}
\end{equation}
The symmetry relation $U(r)=U(-r)$ implies $j_{+}(F)=j_{-}(-F)$, and
$J_{nuc}$ is an even function of $F$ while $J_ {trans}$ is an
odd function of $F$. We find
\begin{equation}
j_{+}=D \frac{\tilde A _{1} F}{kT} \left( 1-\exp (-Fs/kT) \right)^{-1}
\label{ratef}
\end{equation}
where $\tilde A_{1}$ is the normalization constant
given by Eq. (\ref{normconstant}) with a renormalized frequency and
an energy of the new minimum
\begin{eqnarray}
\tilde \omega _{0} =\omega _{0}\sqrt{1+\frac{U^{(4)}}
{2m^{3}\omega_{0}^{6}} F^{2} }
\;\; , \label{omegare} \\
\tilde E _{0} =E _{0}\left(1+\frac{F^{2}}{2m\omega_{0}^{2}E_{0}} \right)
\;\; .
\label{energyre}
\end{eqnarray}
In Eq. (\ref{omegare}) we defined $U^{(4)}\equiv \partial _{r}^{4}U|_{0}$.
The nucleation current $J_{nuc}$ and the total current $J_{trans}$ can then
be written
in the form
\begin{eqnarray}
J_{nuc}=\frac{F\tilde A_{1}}{m\gamma} \coth \left(\frac{Fs}{2kT}\right)
\;\; , \label{jfnuc} \\
J_{trans}= \frac{F\tilde A_{1}}{m \gamma }\;\;,
\label{jftot}
\end{eqnarray}
respectively. This yields a relation between the two currents,
$J_{trans} = J_{nuc}\tanh(Fs/2kT)$. A similar result has been derived in
Ref.
\cite{HAB73} for photo-generated currents. According to our assumptions,
the results are valid in leading order with respect to $F$.
An expansion gives
\begin{equation}
J_{trans} = \frac{Fs}{2kT} \; J_{nuc} \;\;.
\label{HABER}
\end{equation}
Since the nucleation rate is proportional
to $s^{-1}$, the total current is independent of $s$ as it must be.\\
\indent
We now present two derivations to show that the transport current
Eq. (\ref{trans}) is indeed just determined by the imbalance
of the forward and backward nucleation rates.
For weak driving forces the transport current is
equal to the density of free carriers, $n_{\rm free}$, multiplied by their
drift velocity $u = \mu F$,
\begin{equation}
J_{trans} = u n_{\rm free} = \mu F n_{\rm free} .
\label{2trans}
\end{equation}
As usual, the mobility and the diffusion constant are related via
the Einstein relation $D=\mu kT$.
The number of free carriers is determined by
the balance of the generation of free particles and their
recapture into the well, $J_{nuc} = N_{\rm free}/\tau$ with
$N_{\rm free} = L n_{\rm free}$. Here, $J_{nuc}$ determines the
frequency with which free particles are generated, and
$N_{\rm well}\approx 1$. The lifetime $\tau$ which determines the
recapture back into the well is a function of $F$.
But to leading order in $F$, the equilibrium lifetime
Eq. (\ref{tau1}) is all we need.
Eliminating the density of free carriers and using the Einstein relation,
we find, $J_{trans} =  \mu F \tau J_{nuc}/L= (Fs/2kT) J_{nuc} $.
This is in accordance with Eq. (\ref{HABER}), and
thus $J_{trans} = j_{+} - j_{-}$ holds.\\ \indent
A second derivation of this result proceeds as follows.
We still consider a periodic potential with period
$L$ and which is symmetric around the origin.
It is sufficient to consider the range $0 \le r <L$.
We view the transport current as a consequence of the
source currents $j_{+}$ at $r = s$ and $j_{-}$ at
$r = L -s $. The wells act as particle absorbers.
The solution to this problem is a superposition of solutions
to two problems each with one source alone.
The current which flows back from $r =s$ in the presence of the
source $j_{+}$ is denoted by $j_{+}^{L}$ and the current which flows
forward to the next well by  $j_{+}^{R}$. Similarly, the source
$j_{-}$ alone leads to a current back into the well at $r = L$
denoted by $j_{-}^{R}$ and a current into the well at $r = 0$
denoted by $j_{-}^{L}$. The transport current is then given by
\begin{equation}
J _{trans} = j_{+}^{R} + j_{-}^{L} .
\label{trans1}
\end{equation}
Continuity of current requires
\begin{equation}
j_{+} = j_{+}^{R} - j_{+}^{L}, \;\;\; j_{-} = j_{-}^{R}
- j_{-}^{L} \;\; .
\label{trans2}
\end{equation}
As we have seen already the source currents have the symmetry
$j_{+} ( - F) = j_{-} (F)$.
The homogenous recombination currents
$j_{+}^{R}$ and $j_{-}^{L}$ are related by symmetry according to
$j_{+}^{R} (-F) = - j_{-}^{L} (F)$. On the other hand,
the initial recombination currents are even functions of the field
$j_{+}^{L}(-F) = j_{+}^{L}(F)$ and $j_{+}^{R}(-F) = j_{+}^{R}(F)$
and moreover, they are equal in magnitude but differ in their sign,
\begin{equation}
j_{+}^{L}(F) = - j_{-}^{R}(F) \:\; .
\label{trans3}
\end{equation}
The initial
recombination currents are maximal for $F = 0$. With increasing $F$
the recombination current $j_{+}^{L}$ decreases because homogeneous
recombination increases. Similarly, the initial recombination current
$j_{-}^{R}(F)$ decreases because there are fewer carriers activated into
the high energy region of the potential. Using Eq. (\ref{trans1})
and Eq. (\ref{trans2}) we obtain
$J_{trans} = j_{+} - j_{-} + j_{+}^{L}(F) + j_{-}^{R}(F)$.
But as we have seen the sum of the two initial recombination currents
cancel one another and thus $J_{trans} = j_{+} - j_{-}$.
Thus at low fields the difference of the two activation rates
directly determines the transport current. This can also be shown by a
direct
calculation.\\ \indent
The direct relationship of the rates $j_{+}$ and $j_{-}$
to the transport current
demonstrates that these are physically meaningful and useful quantities.

\section{Equilibrium kink nucleation}
\label{kink}
In this section we investigate the dynamics of a string of particles
coupled to each other harmonically and moving in a sinusoidal
potential. The particles are subject to damping and
noise and might be subject to an external driving force.
This model is known as the driven and damped sine-Gordon chain.
It has a long history and due to its wide range of applicability,
from kinks in surface steps on various materials to the motion of fluxons
in long Josephson junctions, has been widely studied
\cite{HTB90,KAW82,Seeg,Loth,SS66,McCum,BBLT,Buet81,Gill,Boch,OGU83,March,Alf}.
In the overdamped limit, which is of interest here,
there are only two types of elementary excitations. There are small
amplitude phonon like excitations and, more interstingly, highly non-linear
structures, called kinks or solitary structures, which describe the
transition from one valley to another. In chemical physics, kinks
are discussed in various polymers \cite{Bernasconi}.
Our concern is the statistical mechanics of such a system
which we take to be so large
that one can define a density of kinks and anti-kinks. Of interest
is a theory of the thermal equilibrium nucleation of kink-antikink
pairs and particularly the role of initial recombination
\cite{BUC95}. We show that a theory of equilibrium
kink nucleation, i.e., for a vanishing external driving force,
{\em must} take into account initial recombination.\\ \indent
At small temperatures the kink-antikink gas is diluted, and the kink
density sets an upper length scale over which nucleation and
annihilation processes have to occur. In the framework of equilibrium
statistical mechanics,
kinks and antikinks are regarded as free particles. Below we shall explain
that at equilibrium the distance between a `mathematically unbounded'
kink-antikink pair is infinitely large. Apparently, this seems to
contradict the
notion of a free kink in a finite system. For reasons of consistency, it is
thus necessary to develop a picture of the nucleation and
annihilation processes which permits essentially free diffusive motion
during the lifetime of a kink.\\ \indent
The nucleation, the dynamics, and the recombination of kinks and
antikinks in space-time is schematically illustrated in Fig. 1 of Ref.
\cite{BUC95}.
Out of equilibrium, in the presence of a strong force (see Fig. 1a of Ref.
\cite{BUC95}),
kink and
antikink are driven apart after a nucleation process (empty
triangles) by the force and annihilate eventually with an
antikink and a kink originating from a different nucleation
process (rectangles). This picture has to be qualitatively modified
in the equilibrium case without force. As shown in Fig. 1b of Ref.
\cite{BUC95},
the diffusive motion of the free kinks gives
them a strongly enhanced probability of initial recombination (closed
loops). Neglecting the closed trajectories
is inconsistent with the experimental definition of free kinks.
Former works \cite{HMS88,March} (see also Ref. \cite{HTB90})
do not describe the initial recombination processes
and count only the negligibly small fraction of extended
trajectories (see Fig. 1b of Ref. \cite{BUC95}) which describe homogeneous
recombination.
Consequently, these works arrive at a much too
low nucleation rate with an activation energy $3E_{k}$ associated
with a kink triple, instead of the pair
energy $2E_{k}$. Hence, these works predict an incomprehensible breaking
of the kink-antikink symmetry. A further consequence is a mean kink
lifetime proportional
to $\exp (2E_{k}/kT)$ \cite{HMS88,March,Alf}, whereas our theory leads to
a kink lifetime
proportional to $\exp (E_{k}/kT)$.\\ \indent
To be specific, let us consider now the overdamped sine-Gordon equation
\cite{SS66}
\begin{equation}
\gamma \partial_{t} \theta=-V_{0}\sin \theta + F +\kappa
\partial_{x}^{2}\theta + \zeta \;\;,
\label{eq1}
\end{equation}
which describes the dynamics of an order-parameter field
(a string of particles)
$\theta (x,t)$ in a periodic potential of amplitude $V_{0}$, and with
a coupling constant $\kappa $. Unless otherwise stated, the force $F$
is set to zero. We assume periodic boundary conditions $\theta ({\cal
L}+x,t)= \theta (x,t)$,
where ${\cal L}$ is the sample length which exceeds every other
relevant length scale of the problem (except the diverging
size of the mathematical critical nucleus). Furthermore, $\zeta $ denotes
a weak white noise force associated with the temperature $T$, i.e. with
zero mean $\langle \zeta \rangle =0$ and the correlation function
$\langle \zeta(x,t) \zeta(\tilde
x,\tilde t)\rangle=2\gamma kT\delta (x-\tilde x)\delta (t-\tilde t)$.
The uniform, stationary, and linearly stable states are
given by $\theta _{s,l}= 2l\pi $ (Peierls valleys) with integer $l$,
and have equal energies. There exists an energy functional $E[\theta ]$
such that Eq. (\ref{eq1}) can be rewritten in the form $\gamma
\partial_{t} \theta=
-\delta E[\theta ]/\delta \theta $. For a weak finite force,
two adjacent Peierls valleys are
separated in function space by a saddle which corresponds to a
kink-antikink pair. A kink $\theta _{k} (x-x_{0})$ centered
at $x_{0}$ connects a Peierls valley $\theta _{s,l}$ with its neighbor
$\theta _{s,l+1}$. An antikink is reversely defined by
$\theta _{a}= \theta _{k} (-x+x_{0})$. A kink-antikink pair at
location $x_{0}$ and with a (not too small) separation $r$ can be
written approximately as $\theta _{N}(x)=$
$\theta _{k} (x-x_{0}+r/2)+ \theta _{k}
(-x+x_{0}+r/2)-2\pi (l+1)$. In the presence of a weak force $F$,
the mathematically exact saddle point
of the energy functional corresponds to a pair with a separation
$\xi = \xi _{0} \ln (V_{0}/F)$ where $\xi _{0} =\sqrt{\kappa /V_{0}}$
is now the kink size. In the equilibrium limit where $F\to 0$,
the separation $\xi $ exceeds the inverse kink density $n^{-1}$
or even the system length $\cal L$. In this case, the mathematically
exact critical nucleus has no physical meaning.
In order to overcome the problem of the large critical nucleus,
we introduce an effective critical nucleus
with a separation $s$ of the kink and the antikink which is larger
than a kink width but much smaller than an average
equilibrium separation $L\equiv n^{-1}$ of kinks. The effective critical
nucleus corresponds to the flat barrier with size $s$, in analogy to
the effective saddle discussed in the previous sections.\\ \indent

\subsection{The nucleation rate}
We derive now the kink nucleation rate per length, $J_{nuc}$.
Note, that due to the periodic array of Peierls valleys, a kink-antikink
and an antikink-kink are equivalent excitations.
If the nucleation rate of the former and the latter are denoted by $j_{+}$
and $j_{-}$, respectively, the rate becomes $J_{nuc}=j_{+}+j_{-}$.
Of course, at equilibrium it holds  $j_{+}=j_{-}$.
The presence of spatial degrees of freedom requires a multi-dimensional
Kramers theory  \cite{LAN69}.
The relevant coordinate is the kink-antikink separation $r$ which can be
associated with a quasi-Goldstone mode, besides the
usual translational Goldstone mode of the pair. The quasi-Goldstone mode is
$\delta \theta _{N} /\delta r = \theta _{k}^{\prime}(x-x_{0}+r/2)/2 +
\theta _{k}^{\prime}(-x+x_{0}+r/2)/2 $ associated with an
infinitesimal variation of $r$. The (orthogonal) Goldstone mode associated
with an infinitesimal displacement $\delta x_{0} $ of the pair is
$\delta \theta _{N}/\delta x_{0} = \theta _{k}^{\prime} (-x+x_{0}+s/2)-
\theta _{k}^{\prime}(x-x_{0}+s/2)$.\\ \indent
The stationary Fokker-Planck equation associated with the Langevin equation
(\ref{eq1}) is solved in an analogous way as in Sect. \ref{flat}
\cite{BUC95}. Degrees of freedom which are transverse to the two Goldstone
modes
can be integrated out by Gaussian integration \cite{Buet81}.
The rate per length of a kink-antikink pair becomes then \cite{BUC95}
\begin{equation}
j_{+}=\frac{1}{L}\frac{\tilde Z_{N}}{\tilde Z_{s}} \;\frac{
\int _{0} ^{\eta _{1}(L)}\: d\eta_{1} }{  \int _{0} ^{\eta_{0}(s)}
\:d\eta_{0}} \exp (-2E_{k}/kT) \;\;.
\label{eq7}
\end{equation}
One concludes an activation energy $E_{0}=2E_{k}$,
where $E_{k}=8\sqrt{\kappa V_{0}}$
is the equilibrium-kink energy \cite{Buet81}. The variables $\eta _{0}$ and
$\eta _{1}$ are the orthonormal-mode
coordinates which belong to the kink-separation mode
and to the translational mode, respectively. It holds
\begin{equation}
 d\eta_{0}^{2} = d r ^{2}
\int (\delta \theta _{N}/\delta r)^{2} dx \;\;\;,
 \;\;\;  d\eta_{1}^{2} = d x_{0}^{2}
\int (\delta \theta _{N}/\delta x_{0})^{2} dx \;\;.
\label{eq9}
\end{equation}
Using the quasi-Goldstone and the Goldstone modes given above,
one finds for the ratio of the integrals in Eq. (\ref{eq7}) a
value $2L/s$. The normalized partition function of the damped
degrees of freedom at the saddle,
\begin{equation}
\frac{\tilde Z_{N}}{\tilde Z_{M}}=\frac{1}{2\pi}
\sqrt{\lambda_{0}^{M}\lambda_{1}^{M}
\prod_{n=2}^{\infty}\frac{\lambda_{n}^{M}}{\lambda_{n}^{N}}}
\label{eq8}
\end{equation}
contains the stability eigenvalues $\lambda _{n} ^{M,N}$ of the metastable
state (index $M$) and the critical nucleus (index $N$) with respect
to perturbations with a decay (or a growth) $\propto \exp (\lambda t)$.
The (quasi-) zero modes are excluded in the products. For a well
separated pair, Eq. (\ref{eq8}) is
the renormalized partition function of a kink-antikink pair without
self-interaction and is given by the square of the single kink
partition function. Hence, $\tilde Z_{N}/\tilde Z_{M}= 4\Gamma/
2\pi $,
where $\Gamma =V_{0}/\gamma $ \cite{Buet81}.
The kink nucleation rate per length finally becomes
\begin{equation}
J_{nuc}=2j_{+}=2\frac{4 \Gamma}{\pi s}  \; \exp (-2E_{k}/kT) \;\;.
\label{eq10}
\end{equation}
As discussed in Sect. (\ref{flat}),
the $1/s$-dependence indicates the strong dependence of the rate
on initial recombination events in one-dimensional systems.

\subsection{Lifetime of kinks}
In the following, we show that the result (\ref{eq10}) is consistent
with equilibrium statistical mechanics, which predicts an equilibrium
kink density \cite{Buet81}
\begin{equation}
n_{eq}=\sqrt{\frac{2V_{0}E_{k}}{\pi \kappa kT}}\exp (-E_{k}/kT)
\;\;.
\label{eq11}
\end{equation}
In order to derive this result independently, we will calculate first the
kink lifetime $\tau $. By using the balance equation $J_{nuc}\tau = n$ we
obtain
then a kink density which has to be compared with (\ref{eq11}).\\ \indent
The kink lifetime $\tau $ for fixed
antikink density $n$ can be calculated with the help of a
Langevin equation for the kink separation $r$. This Langevin
equation follows from a projection of the sine-Gordon Eq. (\ref{eq1}) onto
the
quasi-Goldstone mode discussed above.
The Fokker-Planck equation which is equivalent
to the Langevin equation, reads in the stationary case
$\partial _{r}( \tilde F P - \tilde D\partial _{r}P)=0$.
Here, we assumed a finite force $F$ for later use. The effective force
acting on the separation coordinate is given by $\tilde F =2 \mu F$ with
a mobility $\mu =  2 \pi \kappa /\gamma E_{k}$, and the
diffusion constant is given by $\tilde D=2 \mu kT / 2 \pi $
\cite{BBLT,Buet81}.
The values of $\tilde F$ and $\tilde D$ for the relative
coordinate $r$ are twice as large as for a single kink. Now, we return to
the case $F=0$. The stationary Fokker-Planck equation must be solved with a
 source
at $r=s$ and with sinks at $r=0$ and $r=n^{-1}$. The source
describes the nucleation of a pair, and the sinks model kink-antikink
annihilation. The mean kink distance $n^{-1}$ corresponds
here to the well distance $L$ introduced in Fig. \ref{fig1}c for one
dimension.\\ \indent
Now we can proceed as in subsection \ref{lfp}.
Integration of the stationary Fokker-Planck equation leads to a
piecewise constant current density, whereas the source implies a
discontinuity of the current density of strength $j_{+}$ at $s$.
The absorbing boundary conditions demand $P(0)=P(n^{-1})=0$. The lifetime
$\tau $ is
defined by the ratio of the total probability $\int ds P$
and the injected current $j_{+}$. One finds then immediately the result
(\ref{tau1}) with $L$ replaced by $1/n$, i.e.
\begin{equation}
 \tau  = \frac{s}{2\tilde D n} \;\; .
\label{kinklifetime}
\end{equation}
The balance equation together
with (\ref{eq10}) implies immediately the equilibrium density
(\ref{eq11}) which proves consistency with statistical mechanics, i.e.,
$n=n_{eq}$.\\ \indent
As it must be, the stationary kink density (\ref{eq11}) is
independent of the specific value of $s$. The $s$
dependence of the rate and of the lifetime can be illustrated in Fig. 1 of
Ref.
\cite{BUC95}.
The kinks which are to be counted in $[0,{\cal L}]$ at a
fixed time $t$
must have been nucleated in a strip $(t-\tau,t)$ of width $\tau \propto s$.
Since the number of counted kinks has to be independent of $s$,
the density of generation events in this strip must be proportional to
$1/s$. A variation of $s$ affects only kinetic quantities like
the time scale, but not
thermodynamic equilibrium quantities like the kink density.

\subsection{Effect of a weak force}
\label{ewf}
In the presence of a weak force $F$, we expect a drift in the order
parameter field $\theta$.
Kinks and antikinks which pass a given location with average
velocity $u$ and $-u$, respectively, each advance the
field by $2\pi$. Thus the average speed of the order parameter field
is
\begin{equation}
\langle \partial _{t}\theta \rangle = 2 \pi u(n+m)\;\; .
\label{general}
\end{equation}
where $m$ and $n$ are the average kink and antikink densities.
In our problem it holds $n  = m$.
To leading order in the field $F$ the velocity is simply
determined by the kink mobility $u = \mu F$ and the kink density
is determined by the equilibrium density.
Thus to linear order in the force $F$
the order parameter has a velocity \cite{Buet81}
\begin{equation}
\langle \partial _{t}\theta \rangle = 4 \pi \mu F n_{eq} \;\; .
\label{displacement}
\end{equation}
We will re-derive this result
directly from the kink nucleation rates.
Like in the single-particle problem (\ref{HABER}) of
Sect. \ref{flat} D the transport current can be directly
related to the asymmetry of the kink
nucleation rates.
The kink nucleation rate $J_{nuc} $
is balanced by kink recombination,
$J_{nuc}  = n/\tau $. Here $\tau $ is the lifetime
of kinks, which as we have seen is at low fields determined
predominately by initial recombination and at $ F = 0 $ is
given by Eq. (\ref{kinklifetime}).
Eliminating the kink density from Eq. (\ref{general}),
we obtain $\langle \partial _{t}\theta \rangle = 4 \pi u \tau J_{nuc}$.
Now we find that the difference of the two
nucleation currents is given by
\begin{equation}
j_{+} - j_{-} = \frac{\tilde Fs}{2\tilde D}J_{nuc}.
\label{JF}
\end{equation}
To leading order in the field, it follows immediately
that the average velocity of the displacement field
is directly determined by the imbalance of the nucleation rates,
\begin{equation}
\langle \partial _{t}\theta \rangle = \frac{2 \pi}{n_{eq}}
(j_{+} - j_{-}) \;\; .
\label{theend}
\end{equation}
This expression is equivalent to Eq. (\ref{displacement}).

\section{Conclusion}
In this paper we have shown that diffusion controlled initial
recombination strongly affects the activation rate in problems
with flat potentials. The effect is particularly important in
one-dimensional
systems where the probability of diffusion back to the original well is
very
large. The absence of a well-defined barrier maximum requires the
definition
of an effective size of the barrier, i.e., a location where
for the evaluation of the rate absorbing
boundary conditions to the probability density are applied.
In one-dimensional systems the time and, as a consequence,
the inverse rate, scale linearly with this size.\\ \indent
Furthermore, we discussed equilibrium kink nucleation as an
example where diffusion controlled initial recombination
determines the kinetics. The effective saddle
corresponds here to an effective critical nucleus. A certain
arbitraryness of the nucleus size reflects the uncertainty
of the notion of a free kink. A specific choice of this size, however,
has to be related to physical considerations: obviously,
it has to be larger than the kink width
but much smaller than the mean kink distance. We have shown
consistency of our results with equilibrium statistical mechanics.
Furthermore, we have shown that the activation rates directly determine
the transport current.
Our work not only demonstrates that rates
and lifetimes are useful quantities even if they depend
explicitly on a length scale which separates free particles
from bound particles. It also demonstrates
that an evaluation of such rates is necessary to provide
a physically meaningful discussion of problems in which
diffusion controlled initial recombination plays a dominant role.


\newpage
\section*{Figure Captions}
\begin{figure}
\narrowtext
\caption{a) Potential well with an activation barrier at $r=s$ and with an
activation energy
$E_{0}$; b) Potential well without a barrier maximum, but with a flat
region; this
case has to be treated by introducing a threshold point $s$. c)
Lattice of potential wells with distance $L$. d) Typical trace of initial
recombination
(ir) and of homogeneous recombination (hr)}
\label{fig1}
\end{figure}


\begin{thebibliography}{9}
\bibitem{BUC95} M. B\"uttiker and T. Christen,
          Phys. Rev. Lett. {\bf 75}, 1895 (1995).
\bibitem{ONS38} L. Onsager,
          Phys. Rev. {\bf 54}, 554 (1938).
\bibitem{CAL83} D. F. Calef and J. M. Deutch,
          Ann. Rev. Phys. Chem. {\bf 34}, 493 (1983).
\bibitem{CAU77} J. L. McCauley Jr.,
          J. Phys. C: Solid State Phys. {\bf 10}, 689 (1977).
\bibitem{DON91} K. J. Donovan, J. W. P. Elkins, and E. G. Wilson,
          J. Phys.: Condens. Mater. {\bf 3}, 2075 (1991).
\bibitem{HAB73} R. Haberkorn and M. E. Michel-Beyerle,
          Chem. Phys. Lett. {\bf 23}, 128 (1973).
\bibitem{HMS88} P. H\"anggi, F. Marchesoni, and P. Sodano,
          Phys. Rev. Lett. {\bf 60}, 2563 (1988).
\bibitem{HM96} P. H\"anggi, F. Marchesoni,
          Phys. Rev. Lett. {\bf 77}, 787 (1996).
\bibitem{BUC96} M. B\"uttiker and T. Christen,
          Phys. Rev. Lett. {\bf 77}, 788 (1996).
\bibitem{KRA40} H. A. Kramers, Physica {\bf 7}, 284 (1940).
\bibitem{HTB90} P. H\"anggi, P. Talkner, and M. Borkovec,
             Rev. Mod. Phys. {\bf 62}, 251 (1990).
\bibitem{MEL91} V. I. Mel`nikov,
          Physics Rep. {\bf 209}, 1 (1991).
\bibitem{MIN87} P. Minnhagen,
          Rev. Mod. Phys. {\bf 59}, 1001 (1987).
\bibitem{KAW82} K. Kawasaki and T. Ohta,
             Physica A {\bf 116}, 573, (1982).
\bibitem{RIS84} H. Risken,
          `The Fokker-Planck Equation' (Springer, Berlin,
          Heidelberg, New York 1984).
\bibitem{BHL83} M. B\"uttiker, E. Harris, and R. Landauer,
          Phys. Rev. B {\bf 83}, 1268 (1983).
\bibitem{BUT88} M. B\"uttiker,
          `Noise in nonlinear dynamical systems: Theory, Experiment,
                 Simulation', F. Moss and P.V.E. McClintock, Eds.
                 (Cambridge University Press, 1989), p.45.
\bibitem{MEL86}  V. I. Mel`nikov and S. V. Meshkov,
                 J. Chem. Phys. {\bf 85}, 1018 (1986).
\bibitem{RIS85} H. Risken and K. Voigtlaender,
          J. Stat. Phys  {\bf 41}, 825 (1985).
\bibitem{RIS88} H. Risken, K. Vogel, and H. D. Vollmer,
          IBM J. Res. Develop.  {\bf 32}, 112 (1988).
\bibitem{POL89} E. Pollak, H. Grabert, P. H\"anggi,
          J. Chem. Phys. {\bf 91}, 4073 (1989).
\bibitem{BUR81} M. A. Burschka and U. M. Titulaer,
          J. Stat. Phys.  {\bf 25}, 569 (1981).
\bibitem{HAR81} S. Harris,
          J. Chem. Phys.  {\bf 75}, 3103 (1981).
\bibitem{LAN69} J. S. Langer,
          Ann. Phys. (N.Y.) {\bf 54}, 258 (1969);
      R. Landauer and J.A. Swanson,
          {Phys. Rev. {\bf 121}, 1668 (1961) };
      H. C. Brinkman, Physica (Utrecht) {\bf 12}, 149
      (1956).
\bibitem{Seeg} A. Seeger,
      Phil. Mag. {\bf 1}, 651, (1956).
\bibitem{Loth} J. Lothe and J. P. Hirth
     Phys. Rev. {\bf 115}, 543 (1959)
\bibitem{SS66}A. Seeger and P. Schiller, in {\em Physical Acoustics,
      Vol. III}, edited by W. P. Mason (Academic, New York, 1966)
\bibitem{McCum} D. E. McCumber and B. I. Halperin
      Phys. Rev. B {\bf 1}, 1054 (1970)
\bibitem{BBLT}  M. B\"uttiker and R. Landauer,
      J. Phys. C {\bf 13}, L325, (1980);
      C. H. Bennett et al.,
      J. Stat. Phys. {\bf 24}, 421, (1981)
\bibitem{Buet81}M. B\"uttiker and R. Landauer,
      Phys. Rev. Lett. {\bf 43}, 1453 (1979);
          Phys. Rev. A {\bf 23}, 1397 (1981)
\bibitem{Gill} V. T. Gillard and W. D. Nix,
     Z. Metallkd. {\bf 84}, 874 (1993);
     Y. M. Huang, J. C. H. Spence, and O. F. Sankey,
     Phys. Rev. Lett. {\bf 74}, 3392 (1995)
\bibitem{Boch} A. I. Bochkarev and Ph. de Forcrand,
      Phys. Rev. Lett. {\bf 63}, 2337 (1989)
\bibitem{OGU83} T. \"O. Ogurtani, Ann. Rev. Mater. Sci. {\bf 13}, 67
     (1983).
\bibitem{March}F. Marchesoni,
          Phys. Rev. Lett. {\bf 73}, 2394 (1994)
\bibitem{Alf}M. Alford, H. Feldman, and M. Gleiser,
      Phys. Rev. Lett. {\bf 68}, 1645 (1992)

\bibitem{Bernasconi} Physics in One Dimension, edited by
      J. Bernasconi and T. Schneider (Springer-Verlag, Heidelberg, New
York,
      1981).

\end{thebibliography}
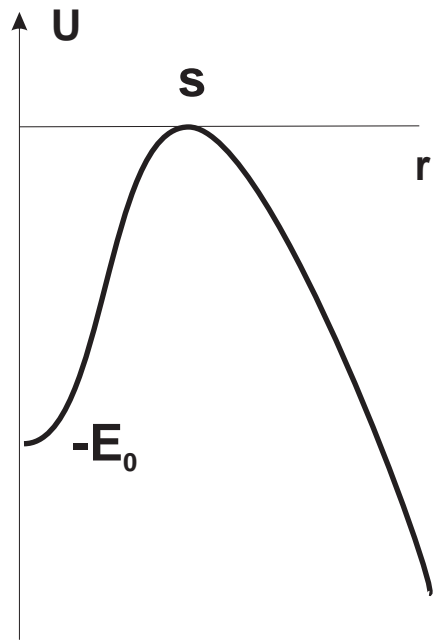
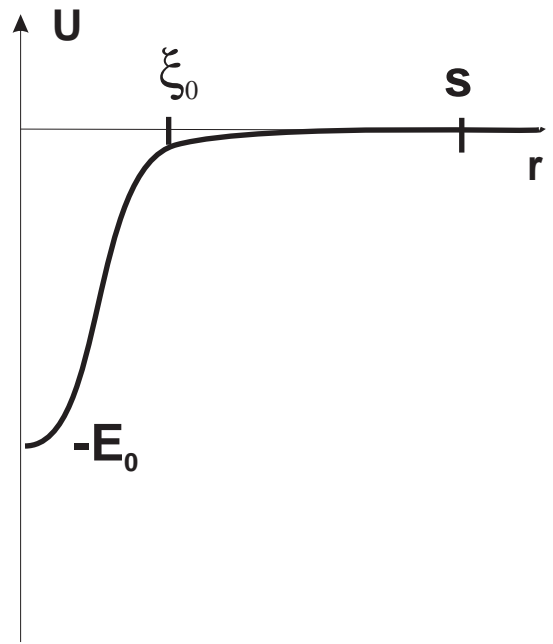
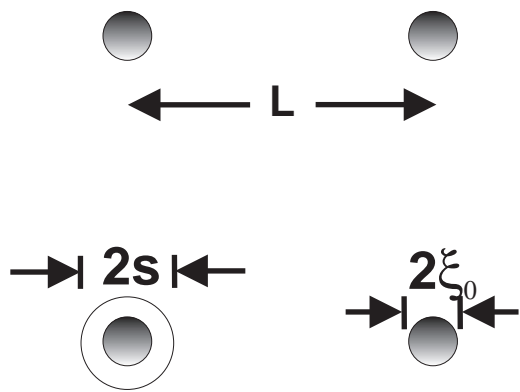
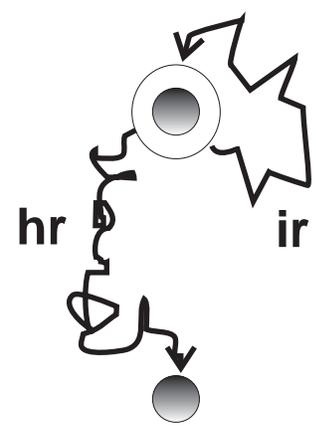
\end{document}